# Gap Filling in the Plant Kingdom—Trait Prediction Using Hierarchical Probabilistic Matrix Factorization

**Hanhuai Shan**[†]                                          SHAN@CS.UMN.EDU
**Jens Kattge**[‡]                                    JKATTGE@BGC-JENA.MPG.DE
**Peter B. Reich**[§]                                        PREICH@UMN.EDU
**Arindam Banerjee**[†]                                   BANERJEE@CS.UMN.EDU
**Franziska Schrodt**[§]                                    FSCHRODT@UMN.EDU
**Markus Reichstein**[‡]                        MREICHSTEIN@BGC-JENA.MPG.DE

[†]Dept. of Computer Science and Engineering, University of Minnesota, Twin Cities, U.S.
[‡]Max Planck Institute for Biogeochemistry, Jena, Germany
[§]Dept. of Forest Resources, University of Minnesota, Twin Cities, U.S.

## Abstract

Plant traits are a key to understanding and predicting the adaptation of ecosystems to environmental changes, which motivates the TRY project aiming at constructing a global database for plant traits and becoming a standard resource for the ecological community. Despite its unprecedented coverage, a large percentage of missing data substantially constrains joint trait analysis. Meanwhile, the trait data is characterized by the hierarchical phylogenetic structure of the plant kingdom. While factorization based matrix completion techniques have been widely used to address the missing data problem, traditional matrix factorization methods are unable to leverage the phylogenetic structure. We propose hierarchical probabilistic matrix factorization (HPMF), which effectively uses hierarchical phylogenetic information for trait prediction. We demonstrate HPMF's high accuracy, effectiveness of incorporating hierarchical structure and ability to capture trait correlation through experiments.

## 1. Introduction

Plant traits are morphological, anatomical, biochemical, physiological or phenological features of individuals or their component organs or tissues, e.g., the height of a mature plant, the mass of a seed or the nitrogen content of leaves (Kattge et al., 2011). They result from adaptive strategies and determine how the primary producers respond to environmental factors, affect other trophic levels, and influence ecosystem functioning (McGill et al., 2006). Plant traits therefore are a key to understanding and predicting the adaptation of ecosystems to ongoing and expected environmental changes (McMahon et al., 2011). To improve the empirical data basis for such projections, in 2007 the TRY project (http://www.try-db.org) was initiated, aimed at bringing together different plant trait databases worldwide. Since then the TRY database has accomplished an unprecedented coverage. It contains 2.88 million trait entries for 750 traits of 1 million plants, representing 70,000 plant species. The consolidated database is likely to become a standard resource for the ecological community and to substantially improve research in quantitative and predictive ecology and global change science.

Despite its large coverage, TRY data are highly sparse, which constrains the usefulness of the joint trait database. Since traits are correlated and they do not vary independently, quite a few quantitative or predictive tasks in ecology require each "referenced" object (It could be an individual plant or a species at a site, but we only use the plant as an example in the following.) to have multiple traits fully available. However, in TRY database, the number of plants with more than same three traits available is extremely small, making it tricky to perform such tasks on TRY data directly. There are two possible solutions: The first is "chopping", i.e., removing all plants with target traits missing. Such a simple strategy results in reduced statistical power and may significantly alter parameter estimates and model selec-tion (Nakagawa & Freckleton, 2008), and for TRY this would actually reduce the data available to a nearly uselessly small number of plants. The second strategy is "filling", i.e., based on non-missing trait entries, filling in the missing entries with predicted values, which yields a complete data set for further processing.

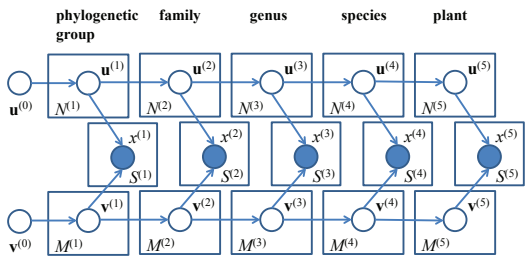

*Figure 1.* Generative process for HPMF.

In this paper, we focus on the "filling" strategy. If we consider the trait data as a plant×trait matrix $X$, with each entry being a trait value, $X$ is a highly sparse matrix and the "filling" problem becomes a matrix completion problem. Meanwhile, the plant kingdom is hierarchically structured. Based on genetic and phenotypic similarity, individual plants can be grouped to species, species to genera, genera to families, and families to phylogenetic groups. Besides trait measurements, each individual plant thus has a set of hierarchically structured phylogenetic information. Therefore, our matrix completion task aims at effectively using such hierarchical structure for better prediction.

Matrix factorizations have achieved great success in matrix completion (Salakhutdinov & Mnih, 2007; Salakhutdinov & Srebro, 2011; Candes et al., 2009). However, one limitation of most methods when applied to TRY data is the inability to use hierarchical phylogenetic information. In this paper, we propose hierarchical probabilistic matrix factorization (HPMF) to incorporate phylogenetic information into matrix factorization for trait prediction. We demonstrate HPMF's high prediction accuracy, effectiveness of incorporating hierarchical structure and ability to capture trait correlation through experiments. Meanwhile, since the phylogenetic tree of plants is constructed under the parsimony principle, plants within a species usually have highly similar trait values, only with small variations due to individual or environmental differences. Therefore, species mean is widely used for gap filling in the ecological community. In experiments, we also show that HPMF generates significantly higher accuracy than prediction using species mean. Although HPMF is proposed specifically for trait data in this paper, in principle it could be generalized and applied to other data with a similar form. In particular, it could be applied to the data matrix with a hierarchy on one side. The hierarchy is balanced such that the distances from the root to all leaf nodes are the same in terms of the levels. In addition, the nodes at the same level have a uniform interpretation, e.g., all nodes at the top level denote phylogenetic groups in our case.

The rest of the paper is organized as follows: Section 2 proposes the HPMF. Section 3 shows experimental results. We give a brief overview of the related work in Section 4 and conclude in Section 5.

## 2. HPMF for Trait Prediction

We propose HPMF to incorporate hierarchical information into matrix factorization for missing value prediction. The hierarchy for trait data includes the phylogenetic group, family, genus, species, and plant from top to bottom. The hierarchy is considered as side information for the plant×trait matrix with missing entries. In addition, we also have data matrices at upper levels, such as species×trait matrix, genus×trait matrix, etc.. These matrices could be constructed from the plant×trait matrix and the hierarchy. The details are discussed in Section 3. We assume the data matrices at all levels to be available before running HPMF.

Denote the data matrix at each level $\ell$ with $X^{(\ell)} \in \mathbb{R}^{N^{(\ell)} \times M}$ for $\ell$ running from the top level 1 to the bottom level $L$. $X^{(\ell)}$ at each level has $S^{(\ell)}$ non-missing entries, and each row $n^{(\ell)}$ and column $m^{(\ell)}$ has a latent factor $\mathbf{u}_n^{(\ell)} \in \mathbb{R}^k$ and $\mathbf{v}_m^{(\ell)} \in \mathbb{R}^k$ respectively[1]. As illustrated in Figure 1, the generative process for HPMF at level $\ell$ is as follows:

1. For each row $n$, generate $\mathbf{u}_n^{(\ell)} \sim \mathcal{N}(\mathbf{u}_{p(n)}^{(\ell-1)}, \sigma_u^2 I)$, where $p(n)$ is the parent node of $n$ in the upper level, e.g., if $n$ is the plant, $p(n)$ is the species of $n$. Note $\mathbf{u}_{p(n)}^{(\ell-1)} = \mathbf{u}^{(0)}$ for $\ell = 1$.

2. For each column $m$, generate $\mathbf{v}_m^{(\ell)} \sim \mathcal{N}(\mathbf{v}_m^{(\ell-1)}, \sigma_v^2 I)$. Note $\mathbf{v}_m^{(\ell-1)} = \mathbf{v}^{(0)}$ for $\ell = 1$.

3. Generate $x_{nm}^{(\ell)} \sim \mathcal{N}(\langle \mathbf{u}_n^{(\ell)}, \mathbf{v}_m^{(\ell)} \rangle, \sigma^2)$ for each non-missing entry, where $\langle \cdot, \cdot \rangle$ is the inner product.

In general, the upper-level latent factor is used as the prior parameter to generate the lower-level latent factor. On the row side, the upper-level latent factor is picked based on the phylogenetic information; on the column side, the one on the same trait is always used.

Denoting latent factor matrices at level $\ell$ with $U^{(\ell)} \in \mathbb{R}^{k \times N^{(\ell)}}$ and $V^{(\ell)} \in \mathbb{R}^{k \times M}$, which has $\mathbf{u}_n^{(\ell)}$ and $\mathbf{v}_m^{(\ell)}$ in columns, the posterior over $\{U^{(\ell)}\}_{\ell=1}^L$ and $\{V^{(\ell)}\}_{\ell=1}^L$ is

$$p\left(\{U^{(\ell)}\}_{\ell=1}^L, \{V^{(\ell)}\}_{\ell=1}^L | \{X^{(\ell)}\}_{\ell=1}^L, \sigma^2, U^{(0)}, V^{(0)}\right)$$

---

[1]When there is no ambiguity, we use $n$ and $m$ instead of $n^{(\ell)}$ and $m^{(\ell)}$ as indexes to avoid clutter.

$$\propto \prod_{\ell=1}^{L} \left\{ \prod_n \mathcal{N}(\mathbf{u}_n^{(\ell)}|\mathbf{u}_{p(n)}^{(\ell-1)}, \sigma_u^2 I) \prod_m \mathcal{N}(\mathbf{v}_m^{(\ell)}|\mathbf{v}_m^{(\ell-1)}, \sigma_v^2 I) \right.$$

$$\left. \prod_{n,m} \delta_{nm}^{(\ell)} \mathcal{N}\left( x_{nm}^{(\ell)}|\langle \mathbf{u}_n^{(\ell)}, \mathbf{v}_m^{(\ell)} \rangle, \sigma^2 \right) \right\}, \quad (1)$$

where $\{\cdot\}_{\ell=1}^{L}$ denotes the data at all $L$ levels ($L = 5$ for TRY data), and $\delta_{nm}^{(\ell)} = 1$ when the entry $(n, m)$ of $X^{(\ell)}$ is non-missing and 0 otherwise. MAP inference on $\{U^{(\ell)}\}_{\ell=1}^{L}$ and $\{V^{(\ell)}\}_{\ell=1}^{L}$ can be done by maximizing the logarithm of the posterior in (1), which boils down to minimizing the regularized squared loss as

$$E = \sum_{\ell=1}^{L} \left\{ \sum_{nm} \delta_{nm}^{(\ell)} \parallel x_{nm}^{(\ell)} - \langle \mathbf{u}_n^{(\ell)}, \mathbf{v}_m^{(\ell)} \rangle \parallel_2^2 \right. \quad (2)$$

$$\left. + \lambda_u \sum_n \parallel \mathbf{u}_n^{(\ell)} - \mathbf{u}_{p(n)}^{(\ell-1)} \parallel_2^2 + \lambda_v \sum_m \parallel \mathbf{v}_m^{(\ell)} - \mathbf{v}_m^{(\ell-1)} \parallel_2^2 \right\},$$

where $\lambda_u = \sigma^2/\sigma_u^2$ and $\lambda_v = \sigma^2/\sigma_v^2$. Next, we explore approaches for doing the MAP inference.

"Diagonally" stacking $X^{(1)}$ to $X^{(L)}$ together yields a matrix $\tilde{X} \in \mathbb{R}^{\tilde{N} \times \tilde{M}}$, where $\tilde{N} = \sum_{\ell=1}^{L} N^{(\ell)}$, and $\tilde{M} = LM$, and $X^{(\ell)}$ is placed from row $\sum_{\ell'=1}^{\ell-1} N_{\ell'} + 1$ to $\sum_{\ell'=1}^{\ell} N_{\ell'}$ and column $M(\ell-1)+1$ to $M\ell$ in $\tilde{X}$. Stacking $U^{(\ell)}$s and $V^{(\ell)}$s together for $\ell = 0 \ldots L$ yields $\tilde{U} \in \mathbb{R}^{k \times (\tilde{N}+1)}$ and $\tilde{V} \in \mathbb{R}^{k \times (\tilde{M}+1)}$ (1 for $\mathbf{u}^{(0)}$ and $\mathbf{v}^{(0)}$). We can construct an undirected graph $W_u$ on the $\tilde{U}$ side, where $W_u(n, n') = 1$ if $n$ is the parent or child of $n'$ based on the phylogenetic information. Similarly, we have $W_v$ on the $\tilde{V}$ side, where $W_v(m, m') = 1$ if $m$ and $m'$ are the same trait at two consecutive levels. Given $W_u$, we can define the graph Laplacian (Luxburg, 2007) $L_u$ and $L_v$ on $\tilde{U}$ and $\tilde{V}$ sides. Therefore, (2) could be rewritten using graph Laplacian for regularization as:

$$E = \sum_{n=1}^{\tilde{N}} \sum_{m=1}^{\tilde{M}} \tilde{\delta}_{nm} \parallel \tilde{x}_{nm} - \langle \tilde{\mathbf{u}}_n, \tilde{\mathbf{v}}_m \rangle \parallel_2^2 \quad (3)$$

$$+ 2\lambda_u \mathrm{tr}\left( \tilde{U} L_u \tilde{U}^T \right) + 2\lambda_v \mathrm{tr}\left( \tilde{V} L_v \tilde{V}^T \right).$$

The problem is not jointly convex on $(\tilde{U}, \tilde{V})$, so one can consider alternately updating $\tilde{U}$ and $\tilde{V}$ to reach a stationary point. Keeping $\tilde{V}$ fixed, the objective is a quadratic form in $\mathrm{vec}(\tilde{U})$, whose solution is given by a linear system of the form $\mathbb{A}\mathrm{vec}(\tilde{U}) = \mathbf{b}$. In spite of the sparsity structure in $\mathbb{A}$, solving such a linear system in every iteration can lead to a prohibitively slow algorithm. Hence, we focus on an efficient alternative: stochastic block co-ordinate descent (Bertsekas, 1999).

At each step, we update $U^{(\ell)}$ or $V^{(\ell)}$ at level $\ell$ through minimizing the objective function in (2) while keeping $U^{(\ell')}$ and $V^{(\ell')}$ at other levels fixed. In particular, at each level $\ell$ ($\ell = 1 \ldots L$), the objective function containing $U^{(\ell)}$ and $V^{(\ell)}$ is given by

$$E^{(\ell)} = \sum_{n,m} \delta_{nm}^{(\ell)} \parallel x_{nm} - \langle \mathbf{u}_n^{(\ell)}, \mathbf{v}_m^{(\ell)} \rangle \parallel_2^2 \quad (4)$$

$$+ \lambda_u \sum_n \left( \parallel \mathbf{u}_n^{(\ell)} - \mathbf{u}_{p(n)}^{(\ell-1)} \parallel_2^2 + 1_{(\ell < L)} \sum_{n' \in c(n)} \parallel \mathbf{u}_n^{(\ell)} - \mathbf{u}_{n'}^{(\ell+1)} \parallel_2^2 \right)$$

$$+ \lambda_v \sum_m \left( \parallel \mathbf{v}_m^{(\ell)} - \mathbf{v}_m^{(\ell-1)} \parallel_2^2 + 1_{(\ell < L)} \parallel \mathbf{v}_m^{(\ell)} - \mathbf{v}_m^{(\ell+1)} \parallel_2^2 \right),$$

where $c(n)$ is the set of child nodes of $n$, e.g, if $n$ is a species, $c(n)$ denotes plants of that species, and $1_{(\ell < L)}$ is an indicator function taking value 1 when $\ell < L$ and 0 otherwise. The regularization terms $\|\mathbf{u}_n^{(\ell)} - \mathbf{u}_{p(n)}^{(\ell-1)}\|_2^2$ and $\|\mathbf{v}_m^{(\ell)} - \mathbf{v}_m^{(\ell-1)}\|_2^2$ keep $\mathbf{u}_n^{(\ell)}$ and $\mathbf{v}_m^{(\ell)}$ close to the corresponding latent factors at level $\ell-1$, and the regularization terms $\sum_{c(n)} \|\mathbf{u}_n^{(\ell)} - \mathbf{u}_{c(n)}^{(\ell+1)}\|_2^2$ and $\|\mathbf{v}_m^{(\ell)} - \mathbf{v}_m^{(\ell+1)}\|_2^2$ keep $\mathbf{u}_m^{(\ell)}$ and $\mathbf{v}_m^{(\ell)}$ close to the corresponding latent factors at level $\ell + 1$ (if applicable). Stochastic gradient descent (SGD) is used to optimize (4) by taking one entry $x_{nm}^{(\ell)}$ at a time and update $\mathbf{u}^{(\ell)}$ and $\mathbf{v}^{(\ell)}$ correspondingly.

An alternative objective function to (2) is

$$E = \sum_{nm} \delta_{nm}^{(L)} \parallel x_{nm}^{(L)} - \langle \mathbf{u}_n^{(L)}, \mathbf{v}_m^{(L)} \rangle \parallel_2^2 \quad (5)$$

$$+ \sum_{\ell=1}^{L} \left\{ \lambda_u \sum_n \parallel \mathbf{u}_n^{(\ell)} - \mathbf{u}_{p(n)}^{(\ell-1)} \parallel_2^2 + \lambda_v \sum_m \parallel \mathbf{v}_m^{(\ell)} - \mathbf{v}_m^{(\ell-1)} \parallel_2^2 \right\},$$

In this case, we do not have $X^{(\ell)}$ for $\ell < L$, so $\mathbf{u}_n^{(\ell)}$ and $\mathbf{v}_m^{(\ell)}$ at levels $\ell < L$ are only used for regularization, we hence refer to it as hierarchy-regularized PMF (HRPMF). However, since there is no data corresponding to latent factors $\mathbf{u}_n^{(\ell)}$ and $\mathbf{v}_m^{(\ell)}$ for $\ell < L$, the minimizer of (5) will have $\parallel \mathbf{u}_n^{(\ell)} - \mathbf{u}_{p(n)}^{(\ell-1)} \parallel_2^2 = 0$ and $\parallel \mathbf{v}_m^{(\ell)} - \mathbf{v}_m^{(\ell-1)} \parallel_2^2 = 0$ for $\ell = 3 \ldots L - 1$. In other words, the latent factors $\mathbf{u}_n^{(\ell)}$ ($\mathbf{v}_m^{(\ell)}$) at intermediate levels $\ell = 2 \ldots L - 1$ effectively reduce to one variable $\mathbf{u}_n^{(L-1)}$ ($\mathbf{v}_m^{(L-1)}$), which does not make use of the hierarchy. In Section 3, we show that HRPMF does not generate satisfactory results.

Once we have $U^{(L)}$ and $V^{(L)}$ inferred, any missing entry $(n, m)$ in the original matrix $X^{(L)}$ can be predicted as $\hat{x}_{nm} = \langle \mathbf{u}_n^{(L)}, \mathbf{v}_m^{(L)} \rangle$.

## 3. Experimental Result

In this section, we present experimental results for trait prediction on TRY data. We show results in prediction accuracy and correlation between traits.

| ID | Name | #Entries | Definition |
|----|------|----------|------------|
| 1 | Specific leaf area (SLA) | 51,848 | The one sided area of a fresh leaf divided by its oven-dry mass |
| 2 | Plant height | 49,595 | Shortest distance of main photosynthetic tissue or reproduction unit on plant and ground level |
| 3 | Dry mass | 96,418 | Dry mass of a whole single seed |
| 4 | Leaf dry matter content (LDMC) | 21,609 | Leaf dry mass per unit of leaf fresh mass (hydrated) |
| 5 | Stem specific density (SSD) | 28,571 | The oven-dry mass of a section of a plant's main stem divided by the its volume when fresh |
| 6 | Leaf area | 52,266 | The one-sided projected surface area of a single leaf or leaf lamina |
| 7 | Leaf nitrogen (LeafN) | 42,760 | Total amount of nitrogen per unit of leaf dry mass |
| 8 | Leaf phosphorus (LeafP) | 20,549 | Total amount of phosphorus per unit of leaf dry mass |
| 9 | Stem conduit density | 3,519 | Number of conduits (vessels and tracheids) per unit of stem cross section |
| 10 | Seed number per reproduction unit | 5,547 | Number of seeds per reproduction unit |
| 11 | Wood vessel element length | 1,019 | Length of a vessel element |
| 12 | Leaf nitrogen content per area | 14,252 | Total amount of nitrogen per unit of leaf area (measured one-sided) |
| 13 | Leaf fresh mass | 12,131 | Fresh mass of a whole leaf |
| 14 | Leaf nitrogen phosphorus ratio (LeafN/P) | 12,712 | Ratio of leaf total nitrogen content versus leaf total phosphorus content |
| 15 | Leaf carbon content per dry mass | 11,562 | Total amount of carbon per unit of leaf dry mass |
| 16 | Seed length | 4,647 | Length of a whole single seed |
| 17 | Dispersal unit length | 3,021 | Length of a whole dispersal unit (seed or fruit) |

*Table 1.* Trait ID, name, number of non-missing entries and definition

### 3.1. Dataset

Currently, data collection and cleaning is still going on in the TRY project. In our experiment, we use a cleaned subset, which is a matrix containing 273,777 plants and 17 traits (Table 1), and 95.3% of entries are missing. The percentage of missing entries in each trait is highly unbalanced, ranging from 66.8% to 99.6%. Starting from the top of the phylogenetic hierarchy, there are 8 phylogenetic groups, 450 families, 7160 genera, 45,824 species, and 273,777 plants.

Given the original plant×trait matrix and the hierarchy, we construct a data matrix on each level of the hierarchy by taking the means. For example, given the plant×trait matrix, together with species for each plant, we can construct a species×trait matrix by taking the mean of the plants in the same species. Similarly, we can also construct a genus×trait matrix, family×trait matrix, etc..

The traits in question have log-normal distribution (Kattge et al., 2011), so we first take the logarithm for entries in the plant×trait matrix, and then calculate the $z$-score for each trait, i.e., for $x_{nm}$ corresponding to plant $n$ and trait $m$, we convert it to $x'_{nm} = (\log(x_{nm}) - lm_m)/ls_m$, where $lm_m$ and $ls_m$ are the mean and standard deviation of the logarithm of trait $m$. After this step, most of the traits are distributed normally ranging from -4 to 4. The results we show are in the transformed space.

### 3.2. Accuracy in Trait Prediction

We first show the accuracy of trait prediction for HPMF by comparing it with other methods.

#### 3.2.1. Algorithms

We run five algorithms on the TRY data: MEAN, PMF, PMF using hierarchical information for initialization level by level (LPMF), HPMF and HRPMF.

The details of each approach are as follows:

**MEAN**: A "hierarchical mean" strategy is used. For example, to predict trait $m$ of plant $n$, among all plants with trait $m$ available for training, if there are plants in the same species as plant $n$, we use species mean for prediction; otherwise, if there are plants in the same genus with plant $n$, we use the genus mean, and so on. In general, among species mean, genus mean, family mean, and phylogenetic group mean, we use the first available one at the lowest level. In the ecological community, taking species mean is a common way to deal with missing data and is highly accurate, since most of the variation is between species and only little within species (Kattge et al., 2011).

**LPMF**: We run PMF (Salakhutdinov & Mnih, 2007) on data matrices level by level following a top-down and bottom-up mode iteratively: first from $X^{(1)}$ to $X^{(L)}$ (top-down), and then from $X^{(L)}$ to $X^{(1)}$ (bottom-up), repeated for several times. At each level $\ell$, we have

$$(U^{(\ell)}, V^{(\ell)}) = \text{PMF}(X^{(\ell)}, initU^{(\ell)}, initV^{(\ell)}),$$

where $initU^{(\ell)}$ and $initV^{(\ell)}$ are initializations of $U^{(\ell)}$ and $V^{(\ell)}$. To incorporate the phylogenetic information, we use the following strategy: in top-down mode, at each level $\ell$, we set $initU^{(\ell)}$ and $initV^{(\ell)}$ based on immediate upper level factorization result $U$ and $V$, i.e., $initU_n^{(\ell)} = U_{p(n)}^{(\ell-1)}$ and $initV_m^{(\ell)} = V_m^{(\ell-1)}$; similarly, in bottom-up mode we set $initU^{(\ell)}$ and $initV^{(\ell)}$ based on immediate lower level factorization result $U$ and $V$, i.e., $initU_n^{(\ell)} = \sum_{n' \in c(n)} U_{n'}^{(\ell+1)}/|c(n)|$ and $initV_m^{(\ell)} = V_m^{(\ell+1)}$, where $|\cdot|$ denotes the number of elements in the set. The intuition is that we use the most updated $U$ and $V$ for the current level initialization immediately.

**HPMF**: We run stochastic block coordinate descent as explained in Section 2. In principle, block coordinate descent allows us to update $\{U^{(\ell)}\}_{\ell=1}^L$ and $\{V^{(\ell)}\}_{\ell=1}^{L}$ in an arbitrary order. Empirically, we do it level by level iteratively following a top-down and bottom-up order. In each iteration, we first do a top-down pass to update $(U^{(1)}, V^{(1)})$ to $(U^{(L)}, V^{(L)})$, followed by a bottom-up pass to update $(U^{(L)}, V^{(L)})$ to $(U^{(1)}, V^{(1)})$, and repeat the process for several iterations. The intuition is that after updating $(U^{(\ell)}, V^{(\ell)})$, we want to immediately use it for regularization in the next level update. Empirically, we observed that such a strategy converges faster than only doing top-down updates repeatedly.

| Phylogenetic info | MEAN | PMF | LPMF | HRPMF | HPMF |
|---|---|---|---|---|---|
| None | 1.0009 ±0.0027 | **0.9743** ±**0.0235** | × | × | × |
| Phylo | 0.9442 ±0.0036 | × | 0.9499 ±0.0058 | 0.9053 ±0.0082 | **0.8812** ±**0.0124** |
| Phylo+family | 0.7699 ±0.0042 | × | 0.7664 ±0.0083 | 0.8013 ±0.0058 | **0.7040** ±**0.0074** |
| Phylo+family +genus | 0.6391 ±0.0036 | × | 0.5960 ±0.0095 | 0.7407 ±0.0335 | **0.5686** ±**0.0063** |
| Phylo+family +genus+species | 0.5703 ±0.0036 | × | 0.4638 ±0.0044 | 0.6999 ±0.0170 | **0.4439** ±**0.0023** |

*Table 2.* RMSE of HPMF and other methods. Latent dimension $k$=15 for matrix factorization methods.

**HRPMF**: We run HRPMF in a similar way as HPMF, but use (5) as the objective function.

**PMF**: We run PMF (Salakhutdinov & Mnih, 2007) directly on the plant×trait matrix without any phylogenetic information, so the prediction is purely based on non-missing traits in the matrix. PMF can be considered as a special case for LPMF and HPMF when no hierarchical information is used.

### 3.2.2. METHODOLOGY

Some plants only have one trait available in our data, and we need at least one trait for each plant to run matrix factorization methods. Therefore, we split the training, test and validation sets as follows: For each plant, if it has at least three traits available, we randomly hold out one trait for test, one trait for validation, and rest for training; if it has two traits available, we randomly hold out one trait for training and one for test; if it only has one trait available, we use it for training. Following such a strategy, each plant has at least one trait in the training set. The test set is used for test and the validation set is used during the training process for early stopping, i.e., after 5 iterations, if the performance on validation set decreases, we stop training. We repeat the holding-out process 5 times to get 5 randomly split datasets, and construct the upper-level matrices for training and validation, but the test set is only at the plant×trait level.

To investigate how the algorithms react to increasing hierarchical information, we first only use the phylo-genetic group information, and gradually add family, genus, and species information one level at a time, until we incorporate all levels into the algorithm.

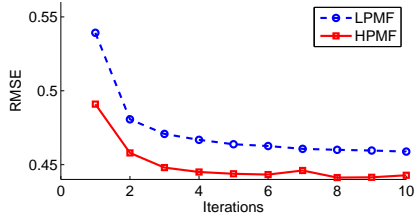

*Figure 2.* RMSE of LPMF and HPMF with increasing number of iterations.

RMSE is used for evaluation. Assuming there are totally $T$ entries for test, $a_t$ is the true value and $\hat{a}_t$ is the predicted value, RMSE is defined as $RMSE = \sqrt{\sum_t (a_t - \hat{a}_t)^2 / T}$.

### 3.2.3. RESULTS

The result for different algorithms are in Table 2. The first row shows the result without using phylogenetic information. In this case, MEAN uses the overall mean of all plants for each trait for prediction. The rest rows show the result with increasing phylogenetic information being used. The result for LPMF, HPMF and HRPMF are obtained from 5 top-down and bottom-up passes. The main message is as follows:
**(1)** For MEAN, LPMF, HPMF, and HRPMF, RMSE keeps decreasing when more phylogenetic information gets incorporated.
**(2)** HPMF outperforms MEAN at all levels by a large margin. We run a paired $t$-test and find HPMF is significantly better than MEAN with a $p$-value smaller than $10^{-5}$ at all levels.
**(3)** HPMF outperforms LPMF, and its advantage is more distinct when only coarse-level phylogenetic information is available. One possible reason is as follows: Finer-level phylogenetic data privide more precise information than the coarse-level data. For example, trait values on plant level are usually closer to trait values on the species level than those on the genus level. Therefore, using species information even just for initialization, LPMF yields a fairly good result. However, for coarse-level phylogenetic data, incorporating it through initialization alone may not work well. A model which captures the hierarchical information explicitly, such as HPMF, is more powerful.
**(4)** HRPMF does not perform well. As explained in Section 2, since $\mathbf{u}_n^{(\ell)}$s and $\mathbf{v}_m^{(\ell)}$s in the upper levels are only used in the regularization and there is no data directly associated with them, the hierarchy is not effectively used.
**(5)** Figure 2 shows how RMSE changes with increas-

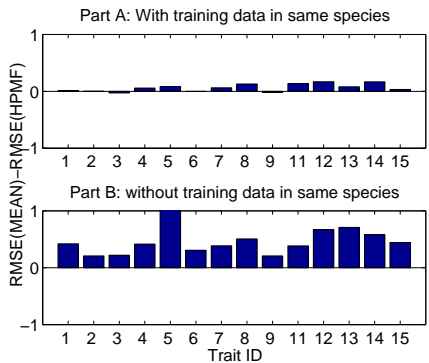

Part A: With training data in same species

Part B: without training data in same species

Trait ID

Figure 3. $RMSE_{MEAN}$-$RMSE_{HPMF}$ on two parts of test data with (Part A) or without (Part B) corresponding training data in the same species available. HPMF performs mostly better than MEAN even on part A, and much better on part B.

ing number of iterations (top-down and bottom-up passes). In the first three iterations, for both LPMF and HPMF, RMSE drops quickly, and then it does not change much along the time.

We do a closer comparison between HPMF and MEAN based on one run of these two algorithms. As we have mentioned, using species mean is highly accurate due to the small variance within the species. Therefore, it is interesting to differentiate the result from the species mean and upper level means for the MEAN strategy, and compare them to HPMF respectively. In particular, the test data is divided to 17 parts, one for each trait, and each part is further divided to two: in part A, for each trait $m$ of plant $n$, there exist plants in the same species with $n$ and their trait $m$ is non-missing. Therefore, the species mean is used for MEAN on this part of data. The rest of the test data are in part B, on which genus or upper-level means are used as available. Such a split of test data does not affect the training process of HPMF, but for test, we compute HPMF's results on part A and B of each trait respectively. The comparison of HPMF and MEAN is presented in Figure 3, where we show $RMSE_{MEAN} - RMSE_{HPMF}$ on each trait, so if the bar is above zero, HPMF performs better, and otherwise MEAN performs better. From Figure 3, we can see that even on part A, where species mean are used for MEAN, HPMF is performing slightly better than MEAN on most traits. The only two traits where MEAN is doing a better job is trait 3 (Seed Mass) and 9 (Stem Conduit Density). These traits have large variation among species, hence species mean is doing well and the collaborative filtering strategy of HPMF is not very helpful by borrowing information from other species. For the result on part B, where upper level means are used for MEAN, the advantage of HPMF is more distinct. This result is important, because it demonstrates HPMF's good prediction performance when trait data in the same species is not available.

### 3.3. True Trait vs Predicted Trait

In the following sections, we will show a variety of results for HPMF other than RMSE.

While RMSE gives an overall accuracy, it masks details about the distribution of values and about fits of predicted to true trait values. Therefore, we do a scatter plot for each trait, using true value and predicted value on test entries. Such scatter plots are useful to the ecological community since they show the distribution of values; the shape, variance and scatter of predictions in relation to the full spectrum of trait values; and whether the predictions are close to the 1:1 line. The results for HPMF are presented in Figure 4, where we use two traits as examples, viz., Wood Vessel Element Length and Leaf Fresh Mass. Each dot denotes one entry in the test set matrix.

The result in the first column is from HPMF without phylogenetic information (i.e., PMF). We show two typical results. For Wood Vessel Element Length, the dots roughly form a horizontal line. For Leaf Fresh Mass, the dots form an "X"-shape plot containing two parts: one part is close to a 1:1 line and the other part is close to a horizontal line. A horizontal line indicates poor prediction, since there is no evident correlation between the true and predicted value, while a 1:1 line indicates good prediction. The reason for the "X"-shape plot is as follows: Leaf Fresh Mass is strongly correlated with Leaf Area (as indicated in Figure 5 in the next subsection), so when we predict the Leaf Fresh Mass, for plants with Leaf Area trait available, PMF is able to use the information on Leaf Area and makes a fair prediction (dots close to the 1:1 line), but for plants whose Leaf Area is not available, PMF cannot do much and gives a poor prediction (dots on the horizontal line).

Starting from the second column, when more phylogenetic information is used, the horizontal line gradually rotates counter clockwise and the shape of the plot becomes more concentrated to the 1:1 line. But note that especially in the second column the slope of the line is still less than 1, suggesting overpredicting at low values and underpredicting at high. Finally, in the last column when all phylogenetic information is used, the plots are roughly 1:1 lines, indicating that the predictions are close to their true values. Thus, for HPMF with all phylogenetic information, one can get accurate prediction in Leaf Fresh Mass even for plants whose Leaf Area is not available, and accurate predic-tion for Wood Vessel Element Length even if there is no trait closely correlated with it.

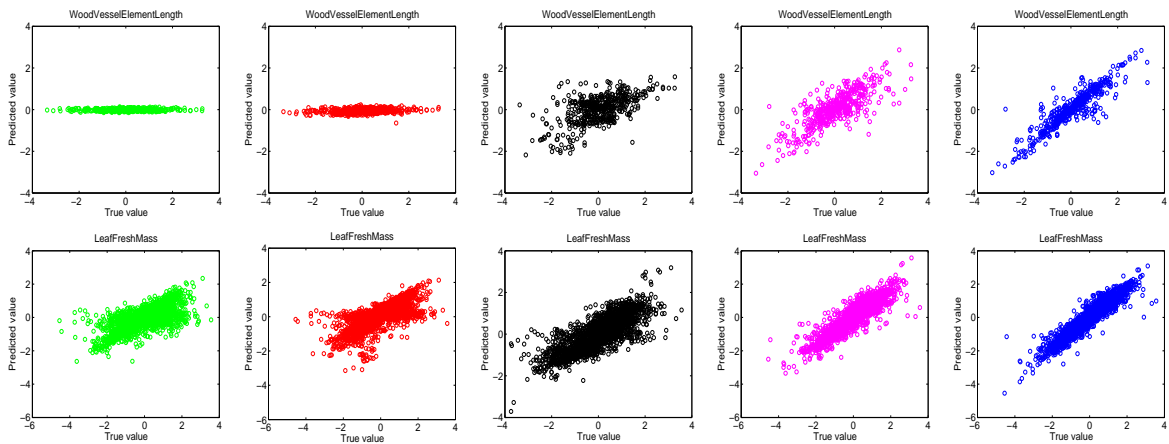

*Figure 4.* Scatter plots for (true value, predicted value) of HPMF on test data of two traits with increasing phylogenetic information. Column 1: no phylogenetic information used. Column 2: phylogenetic group used. Column 3: phylogenetic group and family used. Column 4: phylogenetic group, family and genus used. Column 5: phylogenetic group, family, genus, and species used. When more phylogenetic information is used, the plot becomes closer to a 1:1 line.

### 3.4. Trait Correlation

One of the most interesting aspects for the ecological community is to explore the correlation among the traits and how they vary jointly (Wright et al., 2004; Baraloto et al., 2010). Traits do not vary independently since there are always constraints, e.g., tiny plants cannot have large leaves; and there are also tradeoffs, e.g., plants with thin leaves (low SLA) tend to have high leafN and leafP and thus high photosynthetic rate, and vice-versa. These constraints and trade-offs have implications for ecosystem functioning and the potential of plants to adjust to changing environmental conditions. Therefore, it is interesting to explore the correlation among traits and to further understand the underlying causes and mechanisms. However, due to the high sparsity of the TRY data, it is usually not possible to derive correlation pattern beyond two or three traits, since there are not enough plants having a specific set of more than two or three traits available. Matrix completion technique makes such analysis possible. However, it requires the predicted correlation to be close to the true correlation.

In this subsection, we show the correlation result for pairs of traits. Given the plant×trait matrix, we randomly hold out 80% of entries for training, 10% for validation, and 10% for test. A few test entries do not have training data in the same row (plant) so we cannot predict their values. We ignore these entries and only focus on the ones we can predict.

Given any pair of traits, for plants with both traits available in the test data, we can do a scatter plot for each pair as in Figure 5, where each dot is one plant, and its two coordinates are the two trait values. Figure 5 presents some examples of trait pairs with strong positive or negative correlation. We can see that the correlation from the prediction (green) is close to the correlation from the test set (red). Interestingly, HPMF is picking up accurate correlations, which is a second order information and was not part of the objective function. We explain the observed correlation for each pair as follows:

**LeafFreshMass-LeafArea**: Leaves with large area tend to have higher mass, as leaf thickness is constrained. (physical constraint)

**LDMC-SLA**: Thin leaves with little density (high SLA) tend to have only few structural components to make the leaves robust, they often get their tension for water pressure within the cells (turgor). Thus they have little dry matter content relative to water content: these plants invest little carbon to make the leaves robust. These leaves are often short lived. (leaf economic trade-off (Wright et al., 2004))

**LeafN-LeafP**: For the processes of living, both nitrogen (N) and phosphorus (P) are needed in a specific relation. N is for proteins, and P is for genes and energy distribution within cells. (physiological constraint (Wright et al., 2004))

## 4. Related Work

Apart from using species mean for filling gaps in trait matrices, missing value prediction in ecology is commonly solved by two widely used approaches (Nakagawa & Freckleton, 2008). First is multiple imputation (Rubin, 1987) methods, which typically replace the missing by 3-10 simulated versions. Second is augmentation methods, which implicitly fill in missing entries in the context of model parameter estimation. In these cases the gap filling becomes part of model parameter estimation and the missing entries are not explicitly available.

In the machine learning community, low-rank factorization based algorithms have been developed fast for matrix completion. Salakhutdinov & Mnih (2007) propose probabilistic matrix factorization (PMF). Lawrence & Urtasun (2009) propose a non-linear matrix factorization with Gaussian processes. In addition, there has been work on trace norm regularized matrix factorization (Salakhutdinov & Srebro, 2011), and algorithms which discover low-rank and sparse components (Candes et al., 2009). Recent years have seen emergence of work on incorporating hierarchical structure into matrix factorization. Menon et al. (2011) use the hierarchical structure to help factorize the click through rate matrix on advertisements.

## 5. Conclusions

In this paper, we focus on predicting missing traits for plants in TRY database. We propose HPMF which can incorporate hierarchical phylogenetic information into matrix factorization. We show that HPMF improves the prediction accuracy considerably by effectively using the phylogenetic information in the context of probabilistic matrix factorization. It generates higher prediction accuracy than the species mean strategy, which is considered accurate in the ecological community. We also show that HPMF captures the correlation among the traits accurately.

## Acknowledgments

The research was supported by NSF grants IIS-0812183, IIS-0916750, IIS-1029711, IIS-1017647, and NSF CAREER award IIS-0953274. We thank MPI-BGC Jena, who host TRY, and the international funding networks supporting TRY (IGBP, DIVERSITAS, GLP, NERC, QUEST, FRB and GIS Climate).

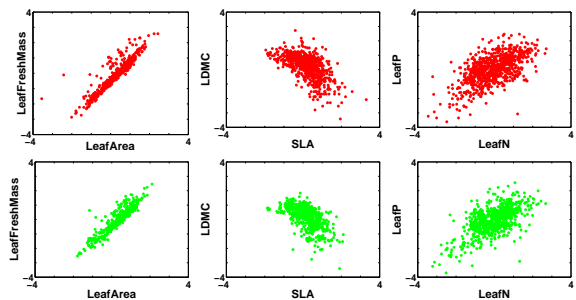

*Figure 5.* Scatter plots for pairs of traits on true test data (first row), and predicted test data (second row). HPMF catches correlation between traits accurately.